\documentclass[12pt]{article}

\usepackage{natbib}
\usepackage{graphicx}
\usepackage{amsmath}
\newcommand{\bsigma}{\mbox{\boldmath$\sigma$\unboldmath}}

\begin{document}

\begin{center}
  {\large\bf Terminus Geometry as Main Control on\\ Outlet Glacier Velocity}
\vspace{2ex}
Martin P.~L\"uthi\\
{Versuchsanstalt f\"ur Wasserbau, Hydrologie und Glaziologie (VAW),
  ETH Z\"urich,
  8092 Z\"urich,
  Switzerland}

\vspace{1ex}
now at
\vspace{1ex}

{University of Zurich, Dept. of Geography, 8057 Z\"urich,
  Switzerland\\
Email: martin.luethi@geo.uzh.ch}
\end{center}


\begin{abstract}
  Ice flow velocities close to the terminus of major outlet glaciers of the
  Greenland Ice Sheet can vary on the time scale of years to hours.  Such flow
  speed variations can be explained as the reaction to changes in terminus
  geometry with help of a 3D full-Stokes ice flow model. Starting from an
  initial steady state geometry, parts of an initially 7 km long floating
  terminus are removed.  Flow velocity increases everywhere up to 4 km
  upstream of the grounding line, and complete removal of the floating
  terminus leads to a doubling of flow speed.  The model results conclusively
  show that the observed velocity variations of outlet glaciers is dominated
  by the terminus geometry.  Since terminus geometry is mainly controlled by
  calving processes and melting under the floating portion, changing ocean
  conditions most probably have triggered the recent geometry and velocity
  variations of Greenland outlet glaciers.
\end{abstract}

\section{Introduction}

Flow velocities close to the terminus of major outlet Glaciers of the
Greenland Ice Sheet can vary substantially on the time scale of years
\citep[e.g.][]{Rignot&Kanagaratnam2006}, and to a lesser degree seasonally
\citep{Joughin&Howat2008}, and episodically during calving events
\citep{Amundson&Truffer2008,Nettles&Larsen2008}.  Several major Greenland Ice
Sheet outlet glaciers accelerated within the last decade
\citep{Joughin&Abdalati2004}, sometimes to double their pre-acceleration
velocity within three years, occasionally followed by a slowdown in the
following years \citep{Howat&Joughin2007}.  The dynamic changes of these
outlet glaciers control the short term evolution of the ice sheet geometry,
and their changing calving flux impacts the ice sheet mass budget, and
therefore sea level.

The close timing of the acceleration of Helheim and Kangerdlussuaq Glaciers on
the east coast \citep{Howat&Joughin2007} and Jakobshavn Isbr\ae\ on the west
coast \citep{Joughin&Abdalati2004} hints to an external forcing not related to
internal dynamic instabilities of the outlet glacier system.  The obvious
possible causes are atmospheric forcing through high meltwater production that
affects basal motion \citep[e.g.][]{Zwally&Abdalati2002}, or the influence of
ocean temperature on terminus melt rate, and therefore the geometry of the
calving front \citep{Holland2008}.  Increased meltwater production can supply
important amounts of water to the ice-bedrock interface by hydro-fracturing
\citep{VanderVeen1998a,Das&Joughin2008}, temporarily increasing the already
very high water pressure under the ice sheet in vicinity of the ice stream
\citep{Luethi&Funk2002a}.  Due to stress transfer from the ice stream trunk to
its surroundings \citep{Truffer&Echelmeyer2003,Luethi&Funk2003b}, stream
velocities are susceptible to changes in basal motion in the surrounding ice
sheet, and could be affected by higher sliding velocities there.  If on the
other hand increased heat flux from the ocean is the driver, thinning of the
floating terminus and higher calving rates would be expected, as were indeed
observed at Jakobshavn Isbr\ae\ \citep{Thomas&Abdalati2003,Holland2008}.

In this contribution we use a three-dimensional full-Stokes ice flow model to
investigate the relation between terminus geometry and ice flow velocity.  The
model results show that the velocity at the grounding line is controlled by
the length of a floating terminus, even in absence of friction at pinning
points.  Removing a floating terminus part by part leads to step-wise
increases in flow velocity, similar to what has been observed at Jakobshavn
Isbr\ae\ \citep{Amundson&Truffer2008} and Helheim Glacier
\citep{Nettles&Larsen2008}.

\section{Flow model}
\label{sec:methods}

The FISMO Full Ice Stream Model was used to investigate the effect of changing
geometry on flow velocities.  The finite element code FISMO solves the Stokes
equations for slow, incompressible flow with variable viscosity, expressed in
terms of the field variables velocity $\mathbf{v}$ and pressure $p$
\begin{subequations}
  \label{eq:stokes}
  \begin{align}
    - \nabla p + \eta\, \nabla^2 \mathbf{v} +
    2\,\mathbf{D} \cdot \nabla \eta + \rho \mathbf{g} &= \mathbf{0}\,,\\
    \mathrm{tr}\mathbf{D} = \nabla \mathbf{v} &= 0\,,
  \end{align}
\end{subequations}
where $\mathbf{D}=\frac{1}{2}\left( \nabla\mathbf{v}+(\nabla\mathbf{v})^T
\right)$ is the Cauchy strain rate tensor.  The viscosity $\eta$ of glacier
ice is strain rate dependent and is calculated according to Glen's flow law (a
power-law rheology) as
\begin{equation}
  \label{eq:glen}
    \mathbf{D} = A\, \tau^{n-1}\, \bsigma^d  \qquad \Longleftrightarrow \qquad
    \eta       = \frac{1}{2}\,A^{-\frac{1}{n}}\,  II_{\mathbf{D}}{}^{\frac{1-n}{n}}\,.
\end{equation}
where $\bsigma^d$ is the deviatoric stress tensor, $II_{\bsigma^d}= \left(
  \frac{1}{2} \sigma^d_{ij}\sigma^d_{ij} \right)^{\frac{1}{2}}$ and
$II_{\mathbf{D}} = \left( \frac{1}{2} D_{ij}D_{ij} \right)^{\frac{1}{2}}$ the
the second invariants of $\bsigma^d$ and $\mathbf{D}$, $A=215
\,\mathrm{MPa}^{-3}\,\mathrm{a}^{-1}$ is the rate factor commonly assumed for
temperate ice \citep{Paterson1999}, and $n=3$.  To avoid inifinte viscosity at
vanishing strain rates, a small constant $\epsilon=1\cdot
10^{-6}\,\mathrm{a}^{-1}$ was added to $II_{\mathbf{D}}$.

Velocity was prescribed as Dirichlet boundary condition on the parts of the
domain representing bedrock.  On the parts of the boundary in contact with air
a zero stress boundary condition was applied (which in the Galerkin finite
element method employed requires no effort).  On the faces in contact with the
ocean, normal stress was set equal to the hydrostatic water pressure $\sigma_n
= -\rho_\mathrm{water} g z$ (ocean level is assumed at $z=0$).  To limit the
geometrical extent of the model, a stress boundary condition was prescribed on
the boundaries to the inland ice, with the normal stress equal to ice
overburden pressure $\sigma_n = -\rho_\mathrm{ice} g (z_s-z)$ (the minus sign
indicates a compressive force).  The latter boundary condition has been tested
to work well for an infinite inclined slab of ice, and is useful since it does
not force ice flow into the computational domain.

Equations (\ref{eq:stokes}) and (\ref{eq:glen}) together with the boundary
conditions were solved numerically with the FISMO finite element (FE) code,
which builds on the Libmesh FE-library (Kirk, 2007) that uses the PETSc
parallel solver library.  To obtain a numerically stable and divergence-free
velocity solution, Q2Q1 isoparametric Taylor-Hood Elements on 27-node
hexahedra were used.  In Libmesh all boundary conditions are enforced with a
penalty method.

The nonlinear equation system arising from Equation (\ref{eq:glen}) was solved
with a fixed-point iteration.  An ALE (arbitrary Lagrange-Euler) formulation
was used in an explicit time stepping scheme for the evolution of the model
geometry.  Given the velocity $\mathbf{v}$ from the previous time step,
vertical mesh node positions $z$ at the free boundaries (surface and floating
terminus) are updated during the time step $\Delta t$ according to
\begin{equation}
  \label{eq:6}
 \Delta z = \left( \left(\mathbf{v}\cdot \mathbf{n}\right)_z + b \right) \Delta t
\end{equation}
where $\mathbf{n}$ is the face normal, index $z$ indicates the vertical
component, and $b$ is the annual net balance, i.e.~the amount of ice added or
removed during a year at the surface or under the floating terminus.  The time
step $\Delta t$ was chosen so that the maximum vertical displacement was
$1\,\mathrm{m}$.  After each time step, all mesh nodes were adjusted to their
initial relative positions between surface and bedrock.

\section{Model setup}
\label{sec:assumptions}

The bedrock is parametrized as a fjord geometry that resembles Jakobshavn
Isbr\ae.  The grounding line location is fixed at $x=0$, ice is grounded in
the domain $x>0$ and floating for $x<0$.  Bedrock elevation $z_b(x,y)$ is
assumed to be $z_b=0$ outside of a straight channel geometry that is
parametrized in the domain $x>0$ by
\begin{align}
  \label{eq:channel}
  z_c(x) &=
  \begin{cases}
      1 + \beta_r(x_{r}-x)\,, & x < x_r \\
      1 \,, & x_r \le x < x_t \\
      e^{-(\beta_{t}(x_t-x))^2}, & x \ge x_t \\
  \end{cases}\notag\\
  z_b(x,y) &= - z_c(x) \, e^{-(\mu y )^2}\,H_{m} 
\end{align}
where $z_c(x)$ is the unit centerline depth of the channel and $\mu$
determines the cross-sectional shape of the channel.  The position $x_t$
determines where the channel reaches is maximum depth and $x_r$ is the
position from where on the channel starts raising towards the calving front
(parameters are given in Table \ref{tab:params}).  The initial ice surface was
parametrized with
\begin{equation}
  \label{eq:surface}
  z_s(x) =
  \begin{cases}
    50 + 4.5 \sqrt{x}\,, & x \ge 0\,,\\
    50 \,, & x < 0\,.
  \end{cases}
\end{equation}
For the initial bottom elevation of the floating terminus the bedrock geometry
given in Equation (\ref{eq:channel}) was extended.  The geometry of the ice
surface and the floating terminus was evolved until it became stationary.
This resulted in surface draw-down in the main ice stream channel where
longitudinal extension rates are highest.

The usually complicated polythermal structure of polar ice streams is
neglected and all ice is assumed to be at the melting point.  This assumption
includes the important effect on ice dynamics of a thick layer of temperate
ice close to the base, the thickness of which is at least $300\,\mathrm{m}$ at
Jakobshavn Isbr\ae\ \citep{Iken&Echelmeyer1993,Luethi&Funk2002a} and might even
amount to $700\,\mathrm{m}$ \citep{Luethi&Fahnestock2008}.  Such an approach
underestimates the stress transfer to the surrounding ice sheet through the
kilometer-thick layer of very cold ice close to the surface
\citep{Truffer&Echelmeyer2003,Luethi&Funk2003b}.

Velocity at the ice-bedrock contact is set to zero and basal motion is
therefore ignored.  This assumption is not realistic, but reasonable for the
task of investigating the importance of geometry change on flow velocities.
Moreover, the flow of Jakobshavn Isbr\ae, for example, seems not to be
dominated by basal motion, and high flow velocities are thought to be largely
due to a thick layer of temperate ice and the steep surface slope.

The geometry was discretized with hexahedra elements (the central part of the
mesh is shown in Figure \ref{fig:modelvelo}).  Since the domain is symmetric
about the $x$-$z$-plane, only one half of the geometry has to be meshed, and
the boundary condition on the $x$-$z$-plane is $v_y=0$.  The computational
mesh of the grounded ice consists of 25x20 elements in the horizontal, and 10
elements in the vertical.  The floating terminus was discretized with up to
30x11x10 elements, depending on terminus length.  Element sizes are reduced
in $x$-direction around the calving front where velocity gradients are
biggest.


The annual net mass balance at the surface is assumed to be elevation
dependent with $b(z)= \gamma (z - z_\mathrm{ELA})$ with a mass balance
gradient $\gamma = 0.005\,\mathrm{a}^{-1}$ and an equilibrium line altitude
$z_\mathrm{ELA}=1100\,\mathrm{m}$, which is the order of measured values close
to the ice sheet margin at Jakobshavn Isbr\ae.  Net mass balance under the
floating terminus was set to $-50\,\mathrm{m}\,\mathrm{a}^{-1}$.

The model experiment was started with an initial geometry as described above,
and with a floating terminus of $7\,\mathrm{km}$ length.  The elevation of the
surface and the bottom of the floating terminus were allowed to evolve until
they reached a stationary state.

\section{Model results}
\label{sec:results}

\begin{figure}
\noindent \includegraphics[width=30pc]{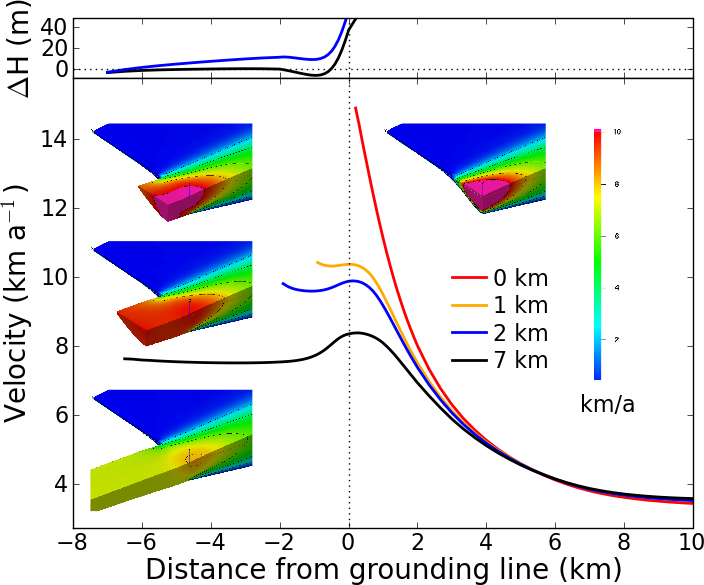}
\caption{Modeled centerline flow velocity at the surface for the terminus
  geometries shown next to the curves.  Decreasing the length of the floating
  terminus affects the value and spatial distribution of flow velocities.
  Grounding line position is indicated with a vertical dotted line.  Top
  panel: surface elevation (blue, displayed $40\,\mathrm{m}$ lower) and
  height above buoyancy (black).}
  \label{fig:modelvelo}
\end{figure}

To investigate the influence of the length of the floating terminus on the
stress state and the flow velocities, parts from the front of the floating
terminus were removed.  New velocity solutions were calculated without any
further evolution of the geometry.  The geometries with terminus lengths of 7,
2, 1 and $0\,\mathrm{km}$ are plotted in Figure \ref{fig:modelvelo} next to the
corresponding centerline velocities.  Flow velocities increase with shorter
terminus length, and nearly double at the grounding line when the floating
terminus is completely removed.

The characteristic local maximum of flow velocity around the grounding line
(black and blue curves in Fig.~\ref{fig:modelvelo}) is due to changing surface
slopes in a small over-deepening plotted in the top panel of Figure
\ref{fig:modelvelo}.  The ice is up to $5\,\mathrm{m}$ below hydrostatic
equilibrium at the grounding line (top plot of Fig.~\ref{fig:modelvelo}) and
reaches equilibrium at about four ice thicknesses along the floating terminus.
A similar surface depression close to the calving front has been found in
other modeling studies \citep[e.g.][]{Lestringant1994}.

\begin{figure}
\noindent \includegraphics[width=20pc]{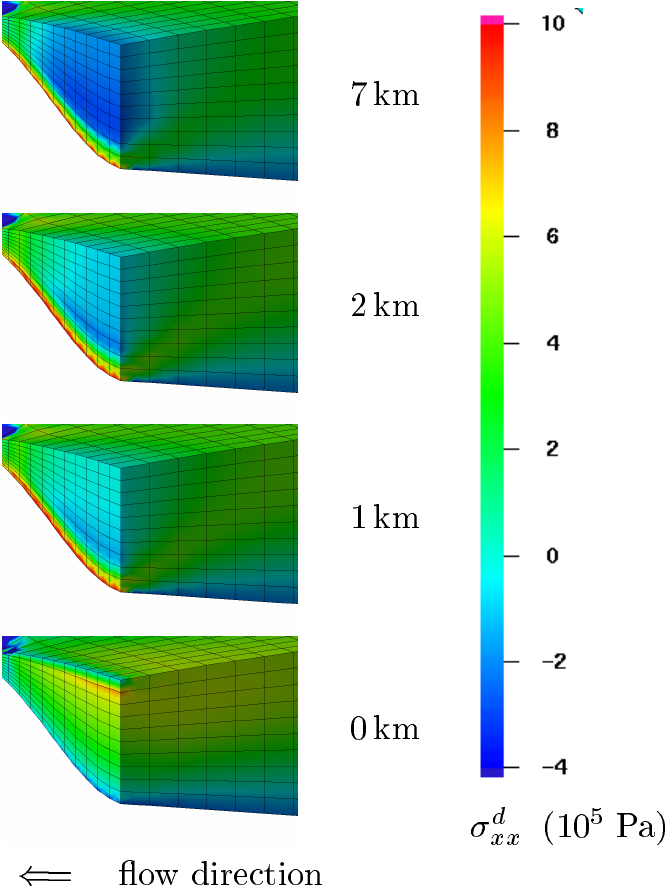}
\caption{Modeled deviatoric stress $\sigma_{xx}^d$ in flow direction at the
  grounding line. The model geometry is sliced along the centerline, and at
  the grounding line (i.e.~the terminus is not shown).  Terminus lengths are
  7, 2, 1 km, and no terminus.}
  \label{fig:modelgeometry}
\end{figure}

\begin{figure}
\noindent \includegraphics[width=30pc]{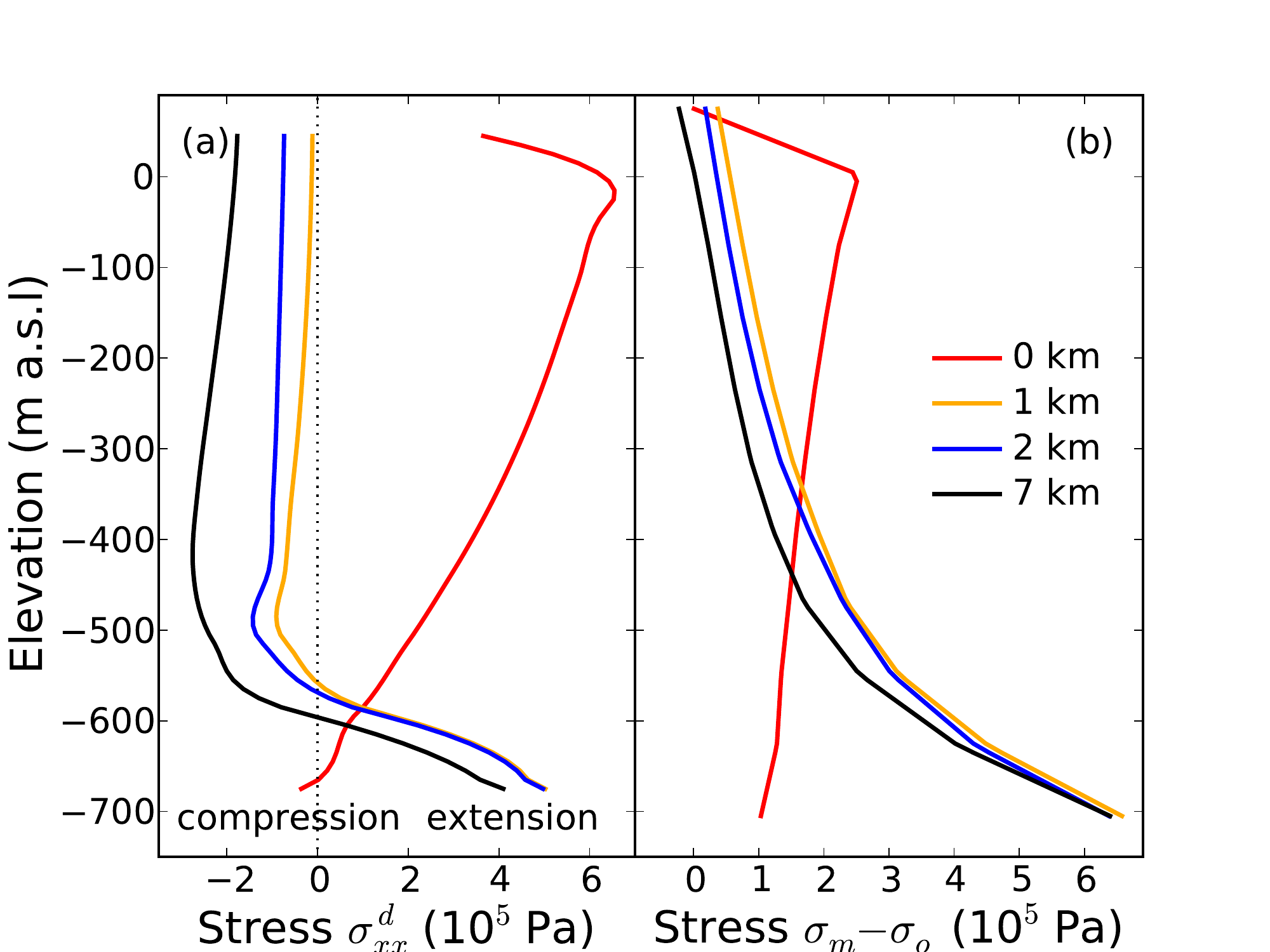}
\caption{(a) Modeled deviatoric component of longitudinal stress
  $\sigma_{xx}^d$ in a vertical profile $20\,\mathrm{m}$ upstream of the grounding
  line.  (b) Deviation of the mean stress $\sigma_m$ with respect to the
  hydrostatic overburden stress $\sigma_o$.}
  \label{fig:sigmaxx}
\end{figure}

The removal of the floating terminus induces changes in the stress state at
the grounding line, as shown in Figures \ref{fig:modelgeometry} and
\ref{fig:sigmaxx}.  Longitudinal deviatoric stress, which determines the rate
of longitudinal extension, changes from negative (compressive) to positive
(extensive) at the grounding line when the floating terminus is removed.  Also
the mean stress (plotted in Fig.~\ref{fig:sigmaxx}b as the deviation from
overburden pressure) shows a marked change with reversed slope and a kink at
the water line (red lines in Fig.~\ref{fig:sigmaxx}).

\section{Discussion}
\label{sec:discussion}

The length and shape of the floating terminus controls the size and
distribution of glacier flow velocity in vicinity of the grounding line, as is
illustrated in Figure \ref{fig:modelvelo}.  There is a striking similarity of
the modeled velocities along the centerline of our prototype outlet glacier
with the measured evolution of terminus velocities at Jakobshavn Isbr\ae\
\citep[Figure 2 in][]{Joughin&Abdalati2004}.  Characteristic features like the
maximum of flow velocity at the grounding line are also visible in the modeled
velocities.  The major difference between model and reality is the measured
increase in flow velocity up to $40\,\mathrm{km}$ upstream of the grounding
line, while our model experiment only shows a speedup in the last
$4\,\mathrm{km}$ upstream of the grounding line.  The difference is due to the
model experiment setup that did not evolve the surfce after removal of the
floating terminus, which corresponds to rapid disintegration of the floating
ice.  The flow acceleration would migrate inland together with the surface
drawdown, as was exemplified in model studies of Pine Island Glacier
(Antarctica) \citep{Schmeltz&Rignot2002,Payne&Vieli2004}.  In these studies
the removal of the terminus immediately influenced ice flow far inland due to
very weak coupling to the bed over large parts of the glacier.  In contrast,
we assumed full coupling, which affects flow velocity only within 5 ice
thicknesses from the grounding line.

Ice shelf buttressing is usually assumed to result from frictional forces
acting on the floating terminus or ice shelf at pinning points, or at its
sides \citep[e.g.][]{Dupont&Alley2005}.  In the absence of friction on the
terminus the total resistive horizontal force in flow direction is that of
water pressure acting as a normal stress on the interface between ice and
ocean.  The integral of this normal stress over the interface area, weighted
by the face normal, is constant and independent of the length and shape,
presence or absence of the terminus.  In the presence of a floating terminus
the horizontal compressive force is evenly distributed over the cross section
at the grounding line (Fig.~\ref{fig:sigmaxx}).  In the absence of the
terminus the water pressure acts only below the water line, which leads to
extensive deviatoric stresses with a maximum at the water line, and
corresponding high extensional ice deformation rates there.  The geometry of
compact floating ice in contact with grounded ice has a major influence on the
stress distribution in the grounding line area.  The model results presented
above show that ice shelf buttressing is mainly an effect of the geometry of
floating portion.

Since the flow velocity is sensitive to terminus geometry, a growing floating
terminus leads to slower flow velocities.  This effect might explain the
observation that Jakobshavn Isbr\ae\ slows in winter when a compact floating
portion forms in the terminus area \citep{Joughin&Howat2008}.

The cause for big changes of terminus geometry, and therefore flow velocity,
is most likely the influence of increased heat flux from the ocean causing
increased melt under the floating portion of the terminus.  For Jakobshavn
Isbr\ae\, a strong increase of ocean bottom temperature has been measured, the
influence of which coincides with thinning of the floating terminus and flow
acceleration \citep{Holland2008}.  Alternatively, changes of the calving
process by ponding meltwater in crevasses could explain the disintegration of
floating termini \citep{Benn&Warren2007}, although observations of increase of
such ponding water have not been made for the Greenland outlet glaciers
discussed above.



\section{Conclusions}
\label{sec:conclusions}

The geometry of the terminus area is the dominant control on the velocity
field close to the grounding line of an outlet glacier.  Short term geometry
changes, such as the disintegration of a floating terminus, calving, or the
creation of embayments in grounded ice, greatly affects the flow field close
to the terminus.

A typical calving event at big outlet glaciers such as Jakobshavn Isbr\ae\
removes up to $400\,\mathrm{m}$ of ice from the glacier terminus.  Measurements
have shown that the glacier velocity reacts immediately (within a 15 minute
measurement interval), but no extra movement could be observed during calving
events \citep{Amundson&Truffer2008, Nettles&Larsen2008}.  Such almost
step-wise increase in flow velocity can be reproduced with a 3D flow model,
when parts of a floating terminus are removed.  Short term flow velocity
variations thus are mainly an effect of stress redistribution, which in turn
is controlled by changes of the terminus geometry.

\section*{Acknowledgments}
  Thoughtful comments by Martin Truffer, Martin Funk and two anonymous
  referees have helped improve the clarity of presentation.
  Funding was provided by the Swiss National Science Foundation
  (200021-113503/1) and by NASA's Cryospheric Sciences Program (NNG06GB49G).

%
%
%
%
%
%
%
%
%
%

\begin{table}[bth]
  \caption{Parameters for the ice stream geometry}
  \label{tab:params}
  \centering
  \begin{tabular}{llrl}
 maximum channel depth & $H_m$ & 1600&m b.s.l.\\
location of deepest point & $x_t$ & 15 & km \\
location of bed inflexion & $x_r$      &  7 & km \\
steepness of trough on centerline \hspace{2ex} & $\beta_{t}$ & 2 & $ 10^{-4}$ \\
centerline slope & $\beta_r$   & -8 &  $10^{-5}$ \\
steepness of sides & $\mu$     &  7  & $10^{-4}$ \\
  \end{tabular}
\end{table}

\end{document}